\documentclass[twocolumn,preprintnumbers,amsmath,amssymb]{revtex4}

\usepackage{graphicx}
\usepackage{graphics}
\usepackage{rotating}
\graphicspath{{Figs/}}
\usepackage{dcolumn}
\usepackage{amsfonts,amsmath,amssymb,bm}

\usepackage{multirow}

\begin{document}

\title{\bf {$d^0$ half-metallic ferromagnetism in CaN and CaAs pnictides: An \textit{ab initio} study}}
\author{Seyed~Mojtaba~Rezaei~Sani$^1$}
\author{Omid Khakpour$^2$}
\affiliation{$^1$School of Nano Science, Institute for Research in Fundamental Sciences (IPM), 19395-5531 Tehran, Iran\\
             $^2$Department of Physics, Payame Noor University (PNU), P. O. Box 19395-3697 Tehran, Iran}

\begin{abstract}

Conventional magnetism occurs in systems which contain transition metals or rare earth ions with partially
filled $d$ or $f$ shells. It is theoretically predicted that compounds of groups IA and IIA with IV and V,
in some structural phases, are ferromagnetic half-metals which made them new candidates for spintronics
applications. Employing density functional theory (DFT) we investigate magnetism in binary compounds
CaN and CaAs. Regarding the structure of analogous magnetic materials
and experimental results of CaAs synthesis, we have considered two cubic structures: rocksalt
(RS) and zincblende (ZB), and four hexagonal structures: NiAs, wurtzite (WZ), anti-NiAs, and NaO.
The calculated results show that CaN in cubic, NiAs, and wurtzite structures, and CaAs only in zincblende
phase have ferromagnetic ground states with a magnetic moment of $1\mu _B$. Electronic structure
analysis of these materials indicates that magnetism originates from anion $p$ states. Existence of flat
$p$ bands and consequently high density of states at the Fermi level of magnetic structures gives rise
to Stoner spin splitting and spontaneous ferromagnetism.

\end{abstract}


\maketitle

\section{INTRODUCTION}

Half-metallic ferromagnetism, which involves metallic conductivity only in
one spin channel, renowned to be a key phenomenon in spintronics. 
Its first-principles introduction by de Groot \textit{et al}. \cite{deGroot1983}
raised enthusiasm among the community to explore this property within materials. 
For then, several reports have been published regarding the prediction and/or observation
of a half-metallic electronic structure in different materials including metallic oxides
\cite{Jemeda2001,Lewis1997}, full and half Heusler alloys
\cite{Borca2000,Ambrose2000,Galanakis2002,Fujii1990}, 
and diluted magnetic semiconductors \cite{Akai1998,Ogawa1999,Yao2005}. 
In spite of several half-metallic compound appeared in the literature,
the search for novel ferromagnetic half-metals is an active research area.
In this context, a significant attention is attracted to the transition metal 
pnictides and chalcogenides with zincblende (ZB) structure 
\cite{Sanyal2003,Yao2005-2,Galanakis2003,Xie2003,Xu2002}.
Although, many of transition metal compounds exhibit half-metallic property in the ZB structure,
these materials crystallize in a different crystal structure including hexagonal NiAs or
orthorhombic structures; hence realization of their half-metallic behavior
requires special requirement for synthesis of metastable ZB structure.
It is claimed that the ZB structure MnAs \cite{Kim2006}, CrAs \cite{Mizuguchi2002}, 
and CrSb \cite{Zhao2001} have been successfully synthesized in the form of thin films. 

In 2004 Kusakabe \textit{et al.} \cite{Kusakabe2004} showed that calcium pnictides 
in the ZB structure, in absence of any transition metal element, 
exhibit half-metal ferromagnetism.
These compounds expose an unconventional type of ferromagnetism 
where the spin polarization is in the anion $p$ electrons without any direct
involvement of $d$ orbital, while in the conventional transition metals magnets,
$d$ electrons are responsible for magnetic ordering. 
This $p$ orbital magnetization is intrinsic and is not triggered by 
crystal defects of any type. The spirit of the exchange is neither
double exchange nor $p-d$ exchange. 
It is explained that these compounds belong to classes of $d^0$ ferromagnets 
which have been recently observed experimentally and/or theoretically\cite{Kusakabe2004}.
Ferromagnetism in the compounds with no valance $d$ or $f$ electron 
has challenged our understanding of magnetism. 

Later on, Sieberer \textit{et al.} \cite{Sieberer2006} and 
Volnianska \textit{et al.} \cite{Volnianska2007} studied ferromagnetism in 
the binary compounds of I$^A$/II$^A$-V elements in tetrahedrally and 
octahedrally coordinated structures. 
They argued that half-metallic ferromagnetism originates from the flat band 
magnetism of holes and may occurs in more structures including RS and NiAs.
It was explained that large cell volume, high ionicity and a slight 
hybridization of anion $p$ orbitals and induced $d$ states around
the Fermi level effectively enhance formation of ferromagnetic (FM) state in the system. 

Our specific aims in this work is to apply accurate first-principles electronic calculations
to investigate magnetic properties  and structural stability 
of two members of binary p magnetic compounds (CaN and CaAs)
in six different structures.
In section III we will discuss structural and magnetic properties of CaN and CaAs. Electronic structures 
of these two compounds are described in section IV.

\section{COMPUTATIONAL METHOD}

We performed first-principles calculations based on Density functional theory (DFT) using 
pseudopotential technique and spin-dependent generalized gradient approximation (GGA) 
in the scheme of Perdew, Burke, and Ernzerhof \cite{PhysRevLett.77.3865}
by means of {\sc Quantum ESPRESSO} package \cite{QE-2009}. 
The relativistic effects are taken into account in the scalar limit,
neglecting the spin-orbit coupling which is expected to be small in light atoms like Ca, N, and As.
Using conventional GGA calculations of electronic and magnetic properties 
of materials which do not consist of atoms with inner $d$ or $f$ 
electrons are thought to be accurate and give reliable results. 
We used ultrasoft pseudopotentials\cite{PhysRevB.59.1758}, a kinetic energy cut off of 35 Ryd 
for the plane wave expansion of Kohn-Sham orbitals, and a kinetic energy cut off of 
400 Ry for the Fourier expansion of electron density. 
The Brillouin-zone integrations were performed using a Monkhorst-Pack mesh of 
$12\times12\times12$ k points for cubic structures and equivalent meshes for hexagonal structures.
Atomic relaxation of the structures with internal parameters was accurately
performed to achieve residual forces less than 0.01 mRy/Bohr and energy 
accuracy of better than 0.001 mRy/$fu$ ($formula~unit$).

\begin{center}
\begin{table}[t]
\newcommand{\mlc}{\multicolumn}
\caption{\label{tab:parameters}
 Computed spin polarization energies $\Delta E^{FM-NM}$ (meV/$fu$), 
 lattice parameters (\AA), and bulk moduli (GPa) for CaN and CaAs in six studied structures. 
 For hexagonal structures, the lattice parameter $c/a$ is given below the first lattice 
 parameter. 
}
\begin{ruledtabular}
\begin{tabular}{lcccccc}
          &  RS  &  ZB  & NiAs &  WZ  & AsNi & NaO  \\
\hline                                              \\
          &   \mlc{6}{c}{$\Delta E^{FM-NM}$}        \\
\\%
     CaN  & -102 & -214 & -177 & -124 &  0   &   0  \\
     CaAs & 0    & -110 &  0   &   0  &  0   &   0  \\ 
\\
          &\mlc{6}{c}{Equilibrium lattice parameters} \\
\\%
     CaN  & 5.00 & 5.45 & 3.39 & 4.19 & 3.73 & 6.39 \\
          &      &      & 1.88 & 1.18 & 1.34 & 0.77 \\
     CaAs & 6.00 & 6.73 & 3.88 & 4.11 & 3.82 & 7.90 \\
          &      &      & 2.10 & 2.22 & 2.16 & 0.75 \\
\\          
          &    \mlc{6}{c}{Bulk modulus}             \\
\\%
     CaN  & 78.4 & 55.4 & 79.8 & 62.3 & 86.9 & 92.1 \\
     CaAs & 40.9 & 24.9 & 39.3 & 30.5 & 41.4 & 49.4 \\
\end{tabular}
\end{ruledtabular}
\end{table}
\end{center}

\section{STRUCTURAL AND MAGNETIC PROPERTIES}

In this work, we take into account six different structures for binary
CaN and CaAs compounds; ZB, wurtzite (WZ), rock-salt (RS), hexagonal NiAs, 
hexagonal anti NiAs (AsNi), and NaO structures.
The tetracoordinated cubic ZB structure is important because several binary compound
exhibit half-metallic ferromagnetism in this structure.
The fact that many binary semiconductors crystallize in the ZB structure
increases the importance of study of this structure for binary half-metals,
because these half-metals are expected to be ideal ferromagnetic
sources for spin injection into semiconductors. 
Also the tetracoordinated WZ structure is the hexagonal analogous of the cubic ZB structure 
and hence is included in our study. 
The significant electronegativity differences in Ca-N (2.0 Pauling) and Ca-As (1.0 Pauling) pairs
\cite{periodic-table} shows the significant contribution of ionic bonding in CaN and CaAs and
hence these compounds may favor a higher coordination atomic arrangement 
to reduce their total energy. 
Therefore, the hexacoordinated cubic RS structure is also considered in this work.
Recently Liu \textit{et al.} have claimed that they have grown a layer 
of RS-CaN on Cu(0 0 1) by means of a new unclear self-assembly mechanism \cite{Liu2008}. 
Since some of magnetic transition metal pnictides (II$^{TM}$-V compounds) 
crystallize in the hexagonal NiAs structure, both NiAs and AsNi (anti-NiAs)
structures were considered for CaN and CaAs.
The natural crystal structure of CaAs seems to be a NaO type lattice \cite{Iandelli1973},
which is derived from the AsNi structure, by a $30^\circ$ rotation clockwise around the z axis,
followed by a shift of $[0,0,0.25]$ in the z direction, and finally an increase of the $a$ and $b$
cell parameters by a factor of $\sqrt{3}$. 

The difference between the minimized total energies of the FM  
and non-magnetic (NM) states, $\Delta E^{FM-NM}$, of CaN and CaAs in all
considered structures are presented in table \ref{tab:parameters}.
To find the minimized total energies, we have optimized lattice constants 
of all structures and atomic positions of the low symmetry NaO structure.
The negative values indicate more stability of the FM phase.
It is seen that CaN has a FM ground state and non-vanishing magnetic moment 
in the RS, ZB, NiAs, and WZ structures while the FM behavior of CaAs 
is limited to the only ZB structure. Both compounds are NM in the NaO phase.
Regardless of the crystal structure, in all magnetic ground states, 
the total magnetic moment per chemical formula unit is 1 $\mu_{B}$. 
In contrast to the transition metal pnictides, the small cation spin moment in 
the magnetic structures of CaN and CaAs is parallel to anion moment
which may be explained by the model of covalent polarization \cite{Lalanne1996}.
Furthermore, table \ref{tab:parameters} lists the equilibrium lattice parameter and 
bulk modulus for both compounds.
It is generally seen that magnetic structures, compared with the nonmagnetic systems,
have higher volume and lower bulk modulus.
As it was mentioned in the introduction, a large cell volume enhances spin polarization
of binary $p$ magnetic materials.

\begin{figure}[!t]
\includegraphics*[scale=0.7]{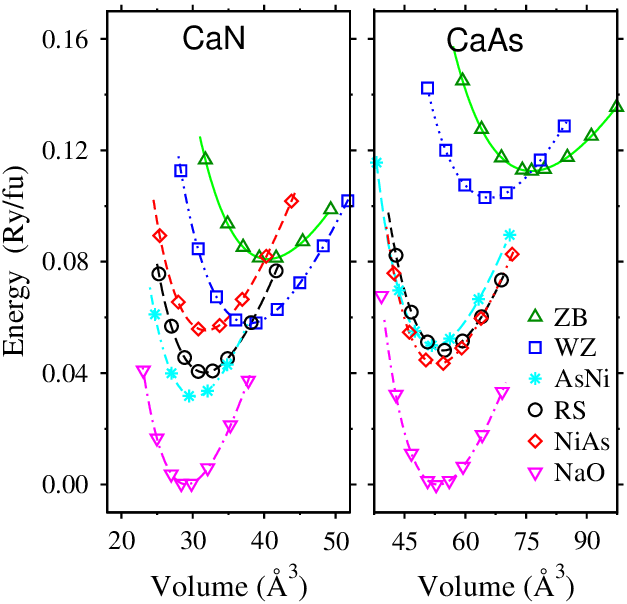}
\caption{\label{fig:E-V}
 Calculated total energy-volume curves of CaN (top) and CaAs (bottom) 
 in different crystal structures.
}
\end{figure}

The calculated total energies in different volumes for CaN and CaAs are plotted 
in figure \ref{fig:E-V}. According to the figure, the structures with hexacoordinated 
anions are energetically more favorable. 
These structures have smaller cell volumes and higher bulk moduli, indicating that their bonds
are stronger due to the smaller interaction distances. 
The ZB structure has the highest total energy and the
experimentally observed structure for CaAs, NaO, with a difference of 
about 0.1 Ry/$fu$ is the most favorable one for both compounds. 
The calculated equilibrium lattice constants of NaO structure for CaAs are 14.93 and
11.15 Bohr for a and c, respectively, which is in good agreement with 
the experimentally found values of 14.85 and 11.19 Bohr. 
The comparison of the ZB and WZ structures shows that the hexagonal arrangement gives rise 
to lower energy for both compounds. 

\begin{figure}
\includegraphics*[scale=0.8]{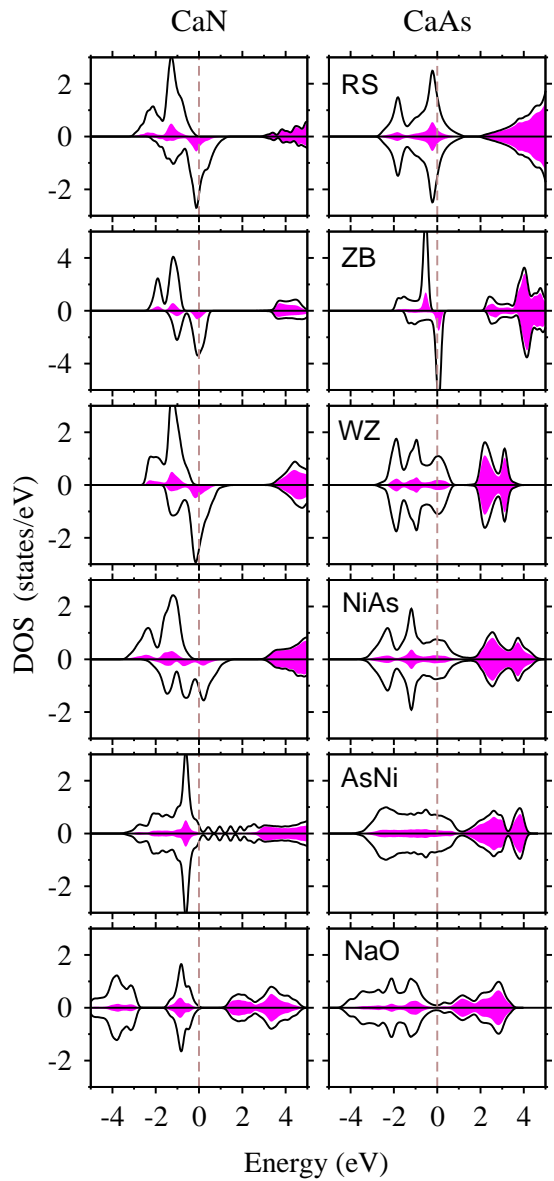}
\caption{\label{fig:DOS} 
  Calculated total density of states (DOS) of CaN and CaAs in six structures. 
  The shaded area shows the contribution of the Ca $d$ orbital an 
  the remained white area belongs to the anion $p$ states.
  Dashed lines show the Fermi energy. Positive and negative values hold for spin up
  and spin down states, respectively.}
\end{figure}

\begin{figure}
\includegraphics*[scale=.9]{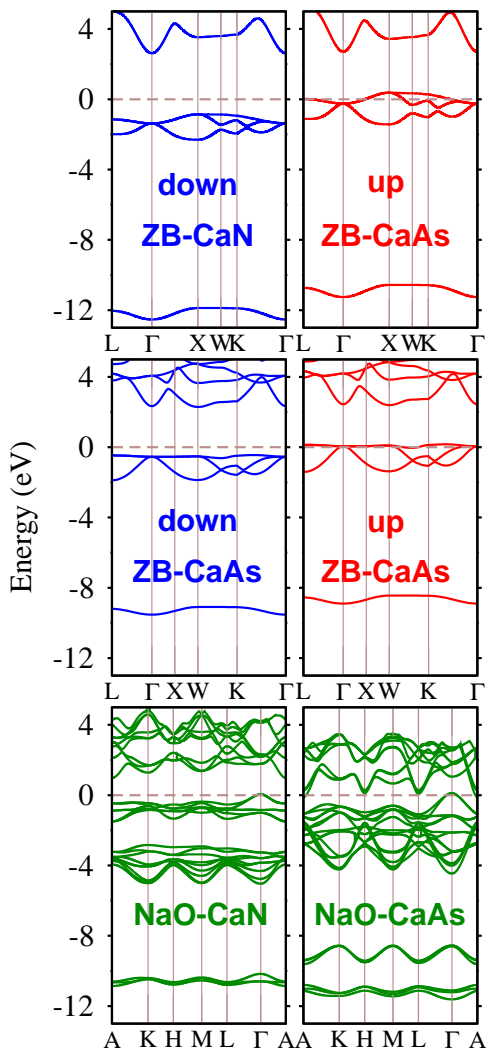}
\caption{\label{fig:band} 
  Obtained band structures of CaN and CaAs in the ZB and NaO structures. 
  The horizontal dashed lines show the Fermi energy. 
  For the ZB structure spin up and spin down band structures are shown 
  on the left and right hand side, respectively.} 
\end{figure}

\section{ELECTRONIC STRUCTURE}

In this section, the detailed investigation of the electronic structure 
of all considered crystal structures is presented, and the essence of ferromagnetism in these
materials is discussed by analyzing their electronic structure. 
According to a previous study \cite{Volnianska2007},
in agreement with ours, for Ca pnictides (CaN, CaP, CaAs, CaSb) in the ZB phase, 
the energy of spin polarization is the largest for CaN and decreases 
with the increase in atomic number of the anion, and almost vanishes for CaSb. 
Therefore, anions should have an important contribution in governing 
magnetization in these compounds. 
For better understanding, we calculated free atom energies in the spin unpolarized and 
polarized states and found that the spin polarization energy of a free N atom (-3.32 eV) 
is significantly higher than a free As atom (-1.58 eV).
This observation may explain stability of FM state in more structures for CaN,
compared with CaAs.

In Fig \ref{fig:DOS}, the spin resolved total density of states (DOS) 
of the systems are presented. 
It is seen that the crystal field splitting between bonding and antibonding states
is more pronounced in the tetra coordinated ZB and WZ structures, compared with other systems.
The magnetic structures including NiAs, WZ, ZB, and RS structure of CaN and 
ZB structure of CaAs obviously exhibit half-metallic property,
as in their majority spin channel there is an energy gap around 
the Fermi level while their minority spin bands cut the Fermi level.
This is different from the half-metallic transition metal pnictides and chalcogenides, 
where their majority spin channel is metallic and the half-metallic gap 
appears in their minority spin channel.
The spin flip gap, the distance between the majority valence band maximum
and the Fermi level, measures the minimum required energy for creating 
a majority spin carrier at the top of the majority valence band.
Hence, systems with larger spin flip gap are expected to be
better source of fully spin polarized current.
Among the studied systems, the largest spin flip gap (0.8 eV) is observed in the ZB-CaN.

It has been argued \cite{Geshi2004} that the half-metallicity in the ZB structure 
is mainly caused by the local bonding environment. 
These alkaline pnictides involve seven valence electrons per formula unit.
Since the anion $s$ states are the lowest energy valence states,
they must be occupied by two electrons, 
and the rest five electron occupy the anion $p$ states which can accommodate six electrons. 
Because of spin splitting, the majority $p$ states lie lower in energy than 
the minority $p$ states, so the majority spin $p$ bands are fulfilled 
by three electrons and the minority spin $p$ bands are partially filled.
As a result of the hole appears in the minority spin valence bands, a total spin moment 
of 1 $\mu_B$ is left on the anion atom. 
But in the case of transition metal pnictides, there are enough valence electrons 
to make the anion $p$ states fulfilled, and the antibonding $t_2-p$ 
bands and the $e$ bands for majority spin channel are also pulled down 
to accommodate the additional valence electrons, 
the Fermi level cut the majority spin bands and a half-metallic gap remains in 
the minority spin channel.
The magnetic moments in these systems are mainly from the transition metals $d$ electrons.

The loss of the half-metallicity can be understood by checking 
the change of the Fermi level position and $p-d$ hybridization with the lattice constants. While the 
lattice constants decrease, the Fermi level will be push up to higher energy and the spin flip gap 
increases, but the $p-d$ hybridization will be strengthened and the valence bands characterized by $p$ 
states are expanded which would reduce the spin flip gap. When the lattice constants are compressed to 
smaller than the critical value, the expanded majority $p$-character bands would cut the Fermi level and
the half-metallicity is destroyed. 

The inter-metallic CaN and CaAs compounds are made of a strong cationic metal and weak anionic 
elements, which are known as the Zintle phase. The Zintle phases stand somewhere in-between metallic and
ionic compounds. In Zintle phases anions do not reach the stable octet state as isolated species,
and hence they need additional bonds to become more stable.
Therefore, in these compounds the anion atoms usually connect covalently together
and form dimers.
Formation of anion dimers in the Zintle phases, electronically balances the system
and leads to closed-shell compounds, i.e. the number of electrons provided by the 
constituting elements equals the number of electrons needed for covalent bonding in the structure.

The spin-dependent band structures along high symmetry lines in 
the Brillouin zone for the best half-metallic structure (ZB)
and the most stable structure (NaO) are given in Fig \ref{fig:band}. 
We know that the anion $p$ states are localized around the nuclei
and hence the valance bands are flat and low dispersed. 
the anion atoms and according to DOS profiles the hybridization between 
the anion $p$ states and the Ca $d$ states is weak. 
reflecting the weak polarized-covalent bonds of the anions
$p$ orbitals and Ca $d$ states.

\section{CONCLUSION}

Conventional magnetism occurs in systems which contain transition metals or rare earth ions with
partially filled $d$ or $f$ shells. Recent experimental and/or theoretical observations of magnetism in
systems with no valance $d$ or $f$ electron challenge our classical understanding of magnetism. It is
theoretically predicted that compounds of groups IA and IIA with IV and V, in some structural
phases are ferromagnetic half-metals which made them new candidates for spintronics applications.
Employing density functional theory (DFT) we investigate magnetism in binary compounds CaN
and CaAs. Regarding the structure of analogous magnetic materials and experimental results of
CaAs synthesis, we have considered two cubic structures: rocksalt (RS) and zinc-blende (ZB), and
four hexagonal structures: NiAs, wurtzite (WZ), anti-NiAs, and NaO.
The calculated results show that CaN in cubic, NiAs, and wurtzite structures, and CaAs only in
zincblende phase have ferromagnetic ground states with a magnetic moment of 1 $\mu_B$ . Electronic structure
analysis of these materials indicates that magnetism originates from anion $p$ states. Existence
of flat $p$ bands and consequently high density of states at the Fermi level of magnetic structures
gives rise to Stoner spin splitting and spontaneous ferromagnetism. Larger exchange interaction in
$2p$ orbitals of nitrogen respect to $4p$ orbitals of arsenic causes that CaN has more structures with
ferromagnetic ground state than CaAs. Slight hybridization with calcium $d$ states increase density
of $p$ states at the Fermi level and enhance ferromagnetism in the system.

\bibliography{references} 
\end{document}